%% file: aggregate.tex
\documentclass[10pt, conference, letterpaper]{IEEEtran}
\makeatletter
\def\ps@headings{%
\def\@oddhead{\mbox{}\scriptsize\rightmark \hfil \thepage}%
\def\@evenhead{\scriptsize\thepage \hfil \leftmark\mbox{}}%
\def\@oddfoot{}%
\def\@evenfoot{}}
\makeatother
\usepackage{graphicx}
\usepackage{times}
\usepackage{graphics}
\usepackage{framed,lipsum}
\usepackage{url}
\usepackage{subfig}
\usepackage{epsfig}
\usepackage{caption}
\usepackage{varwidth}
\usepackage{algorithm}
\usepackage{algorithmicx,algpseudocode}

\newcommand{\CASE}[1]{\STATE \textbf{case} #1\textbf{:} \begin{ALC@g}}
\newcommand{\ENDCASE}{\end{ALC@g}}

\newcommand{\DEFAULT}{\STATE \textbf{default:} \begin{ALC@g}}
\newcommand{\ENDDEFAULT}{\end{ALC@g}}
\newcommand{\DEFAULTLINE}[1]{\STATE \textbf{default:} }
\usepackage{color}
\usepackage{multirow}
\usepackage{multicol}
\usepackage{hhline}
\usepackage{amsthm}
\usepackage{comment}
\usepackage{tabu}

\theoremstyle{plain}

\theoremstyle{definition}

\theoremstyle{remark}

\IEEEoverridecommandlockouts

\begin{document}
\title{Toward Incremental FIB Aggregation with Quick Selections (FAQS)}
\author{
Yaoqing Liu \\liu@clarkson.edu \\
Clarkson University
\and Garegin Grigoryan\\ gg5996@rit.edu  \\Rochester Institute of Technology\IEEEauthorrefmark{1}\thanks{\IEEEauthorrefmark{1}This work was done when Garegin Grigoryan was a student at Clarkson University.} 
}

\maketitle

\pagestyle{empty}

\input{abstract}
\input{intro}
\input{design}

\input{results}

\input{related}


\input{conclusion}
\bibliographystyle{IEEEtran}
\bibliography{net}
\end{document}

%% file: abstract.tex
\begin{abstract}
Several approaches to mitigating the Forwarding Information Base (FIB) overflow problem were developed and software solutions using FIB aggregation are of particular interest. One of the greatest concerns to deploy these algorithms to real networks is their high running time and heavy computational overhead to handle thousands of FIB updates every second. In this work, we manage to use a single tree traversal to implement faster aggregation and update handling algorithm with much lower memory footprint than other existing work. We utilize 6-year realistic IPv4 and IPv6 routing tables from 2011 to 2016 to evaluate the performance of our algorithm with various metrics. To the best of our knowledge, it is the first time that IPv6 FIB aggregation has been performed. Our new solution is 2.53 and 1.75 times as fast as the-state-of-the-art FIB aggregation algorithm for IPv4 and IPv6 FIBs, respectively, while achieving a near-optimal FIB aggregation ratio.

 
\end{abstract}

%% file: intro.tex
\section{Introduction}
\label{sec:intro}

\subsection{FIB scalability problem}
Several factors contribute to the super-linear growth of global Forwarding Information Base (FIB) size. First of all, the tremendous growth of the number of Internet users results in new network prefixes to be allocated and advertised. Second, network operators often divide large block of IP prefixes allocated to an Autonomous System (AS) into smaller ones and advertise them via Border Gateway Protocol (BGP) to enable fine-grained traffic engineering. According to several research studies (\cite{meng2005ipv4}, \cite{cidr_report}), around 50\%  of BGP-announced prefixes are more specific prefixes, i.e., the total address space they cover belongs to large address blocks allocated by Internet Assigned Numbers Authority (IANA). 40\% of these more specific prefixes are attributed to Traffic Engineering, which is used by network administrators to avoid congested paths~\cite{valera201112} or fight against prefix hijacking~\cite{zhao2010routing}. Address fragmentation by multi-homing, a practice to connect an end-user network to more than one network in order to provide high throughput and resilient connectivity, is another source of extra prefixes in a routing table (\cite{zhao2010routing}, \cite{bu2004characterizing}). 
Overall, the number of entries in FIB has increased almost 40 times since 1994, when the current BGP version 4 emerged. In 2017, the size of FIB has approached 710,000 entries for IPv4 and 40,000 for IPv6, and continues to increase with a super-linear pace~\cite{bgp_pic}.


Supporting the current size of FIB and its growth is a challenging task for Default-Free Zone (DFZ) network operators, as they are forced to periodically upgrade their routing hardware in order to fit the FIB into line cards. It is a heavy financial burden for many small Internet Service Providers (ISPs) to migrate old hardware to new one due to the high costs of line cards and operational expenses (\cite{zhao2010routing}, \cite{khare2010evolution}). Some operators avoid upgrading expenses by filtering out specific prefixes with prefix length more than 24, thus affecting the reachability of the Internet~\cite{li2015nexthop}. The increasing size of global FIB may also increase chip space for Ternary Content-Addressable Memory (TCAM) design, the Longest Prefix Match (LPM) lookup time~\cite{Meyer07:RAWS} and energy consumption by line cards~\cite{moradi2015caesar}.

\subsection{Current approaches}
To mitigate the FIB scalability problem, a number of possible solutions were put forward. They can be classified into two broad categories: long-term and short-term solutions. The long-term solutions include revision of the business relations between ASes, e.g., network operators working in the Default-Free Zone (DFZ) can be compensated for keeping all routes in FIB, and re-design of the routing architecture, e.g., splitting address space into a locator (for routing systems) and an identifier (for end systems), may significantly reduce the size of global FIB table, but its wide deployment may take long time~\cite{zhao2010routing}. FIB aggregation falls into the category of short-term solutions. Network operators believe it to be one of the most feasible solutions at this moment as it has a clear benefit and many ISPs are seeking such a solution to reduce their operational costs and mitigate their routing scalability problem~\cite{zhao2010routing}. FIB aggregation does not require changes on routing hardware and routing architecture, and can be applied locally to each individual router. Several FIB aggregation techniques, such as the Optimal Routing Table Constructor (ORTC) algorithm~\cite{draves1999constructing}, can greatly reduce the number of FIB entries for an IPv4 FIB by more than 50\%. When comparing this result to the rates of FIB growth, we infer that the FIB aggregation may prolong a router's lifetime up to 9 years. However, existing FIB aggregation approaches, such as ORTC-based aggregation algorithms, suffer from a number of challenges that remain to be addressed: 

(1) High time costs for processing route updates, including additions, withdrawals and changes. For instance, one of the state-of-the-art FIB aggregation algorithms, \textit{FIFA-S}~\cite{liu2013fifa}, can achieve optimal aggregation ratio for each update, but needs to perform two subtree traversals in the control plane to update an FIB into aggregated and optimal state.  

(2) Individual routing updates result in a significant number of changes in an FIB, called FIB bursts. 

(3) The optimal compression ratio is achieved at the expense of high memory usage: each node generated by the aggregation algorithm in the control plane contains an array of variable size, which stores next-hop candidates to be used for next hop selection for aggregated prefixes.


\subsection{Our contributions}
In this work we introduce a new ultra-fast FIB aggregation algorithm: \textit{FIB Aggregation with Quick Selections (FAQS)}. Different from existing aggregation algorithms, \textit{FAQS} uses a single tree traversal to conduct both FIB aggregation and handle FIB updates. It handles routing updates incrementally, without re-aggregation of the whole forwarding table. On a single BGP update, in the worst case, \textit{FAQS} will traverse only the subtree rooted at the updated node and its parents' nodes. 
Furthermore, unlike \textit{FIFA-S}, \textit{FAQS} keeps only a single next hop at each node and considerably reduces memory consumption for aggregation operations.  The outcome of our improvements is the significant acceleration of FIB aggregation and update handling.
Although \textit{FAQS} is still a heuristic aggregation algorithm, we experimentally proved its superior performance via multiple realistic datasets in different Routing Information Bases (RIBs) with more than 1 billion route updates from Route Views Project~\cite{routeviews} over a 6-year period. 
The results are briefly described as follows:

(1) \textit {FAQS} achieves high and near-optimal compression ratios: reducing the number of FIB entries by up to 73\% for IPv4 and 42\% for IPv6. 

(2) \textit{FAQS} runs up to 2.53 and 1.75 times as fast as existing \textit{FIFA-S} algorithm for IPv4 and IPv6 FIBs, respectively. 


(3) \textit{FAQS} reduces the average number of FIB changes by 30\% for IPv4 routing tables and by 10\% for IPv6 routing tables.

(4)  \textit{FAQS} can save up to 30\% of memory consumption compared with \textit{FIFA-S} algorithm that achieves optimal aggregation ratio. FIFA series~\cite{liu2013fifa} have three algorithms: \textit{FIFA-T}, \textit{FIFA-H}  and \textit{FIFA-S}. We did not compare \textit{FAQS} with \textit{FIFA-T} and \textit{FIFA-H} as both of them are threshold-based aggregation algorithms (not strictly incremental). Namely, when a threshold is reached, both algorithms need to re-aggregate over an entire tree/subtree. However, both \textit{FAQS} and \textit{FIFA-S} are strictly incremental FIB aggregation algorithms for every single update. 

%% file: design.tex
\section{Design}
\label{sec:design}
\subsection{FIB aggregation in a nutshell}
\begin{table}[tbp]
	\centering
	\small
	\caption{FIB aggregation process}
	\captionsetup{justification=centering}
	\label{tab:fibentry}
	\subfloat[Original FIB table]{
		\begin{tabular}{| c | r | c | }
			\hline
			Label & Prefix  & Next Hop   \\ \hline
			A & 141.92.0.0/16 & 1    \\ \hline
			B & 141.92.64.0/18 & 1    \\ \hline
			C & 141.92.0.0/19 & 1    \\ \hline
			D & 141.92.192.0/19 & 2    \\ \hline
			E & 141.92.224.0/19 & 2    \\ \hline
		\end{tabular}
		\label{tab:originalfibentry}
	}
	\\
	\subfloat[Compressed FIB table]{
		\begin{tabular}{| c | r | c |}
			
			\hline
			Label & Prefix  & Next Hop   \\ \hline
			A & 141.92.0.0/16 & 1    \\ \hline
			D & 141.92.192.0/19 & 2    \\ \hline
			E & 141.92.224.0/19 & 2    \\ \hline
		\end{tabular}
		\label{tab:filteredfibentry}
	}
	\vspace{-7mm}
\end{table}

FIB aggregation refers to a process, that merges two or more FIB entries with different prefixes and same next hop into one. While FIB aggregation may significantly compress the size of an FIB, the aggregation process should not change the forwarding behaviors of any packet. Namely, the next hop for any packet should be same before and after aggregation. 
In Table~\ref{tab:originalfibentry}, FIB entries $B$ and $C$ have the same next hop value as the entry $A$, which fully covers IP address blocks of both $B$ and $C$. Hence, excluding the entries $B$ and $C$ from the FIB table will not change the forwarding behaviors of any packets matching against $B$ or $C$, which preserves the \textit{Forwarding Correctness} rule. Excluding the entries $D$ or $E$, in contrast, will not preserve \textit{Forwarding Correctness}, e.g., packets with destination IP addresses from these blocks will be forwarded to the next hop 1 instead of 2. The correctly aggregated FIB is given in Table~\ref{tab:filteredfibentry}.





\subsection{FAQS overview}
\begin{figure}
	\begin{center}
		\captionsetup{justification=centering}
		\begin{minipage}[c]{3in}
			\includegraphics[width=\linewidth]{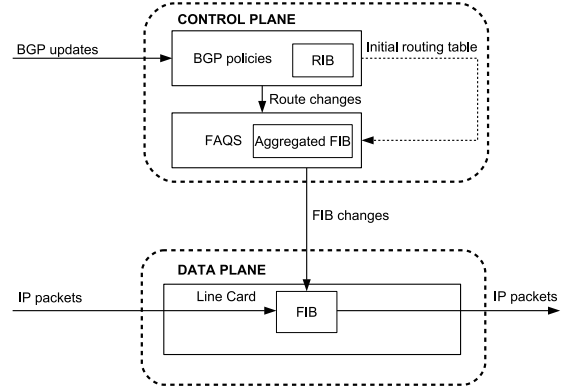}    
		\end{minipage}
		\caption{\label{fig:SAarch}FAQS Module}
	\end{center}
	\vspace{-7mm}
\end{figure}
As illustrated in Figure~\ref{fig:SAarch}, FIB aggregation and \textit{FAQS} algorithm operate in a router's Control Plane, between the RIB and the FIB. When the router boots up, \textit{FAQS} algorithm aggregates the initial set of routes from the RIB and downloads them into the FIB. We call this process \textit{Static FIB Aggregation}. Meanwhile, \textit{FAQS} keeps a copy of the aggregated FIB with various flags to process future route updates. After a routing update, either an addition, a change, or a withdrawal, is advertised via a routing protocol, e.g., BGP, the router first updates its RIB in accordance with BGP decision process. Subsequently, the routing changes are pushed to the aggregation module, where \textit{FAQS} algorithm carries out incremental FIB updates over the aggregated FIB, located in the control plane. A routing update, applied to an aggregated FIB, may not always lead to changes in an FIB. In the meantime, it may result in multiple FIB changes: adding new entries to an FIB, changing next hop values for existing entries or deleting existing entries. If there are FIB changes, \textit{FAQS} installs them in the line cards located in the data plane. 

The remaining part of this section describes both  \textit{Static FIB Aggregation} process and  \textit{Incremental FIB Update Handling} process in detail.

\begin{figure*}[t!]
	\centering
	\captionsetup{justification=centering, width=.28\linewidth}
	
	\subfloat[Initial PATRICIA tree]
	\centering
	\includegraphics[width=.28\linewidth]{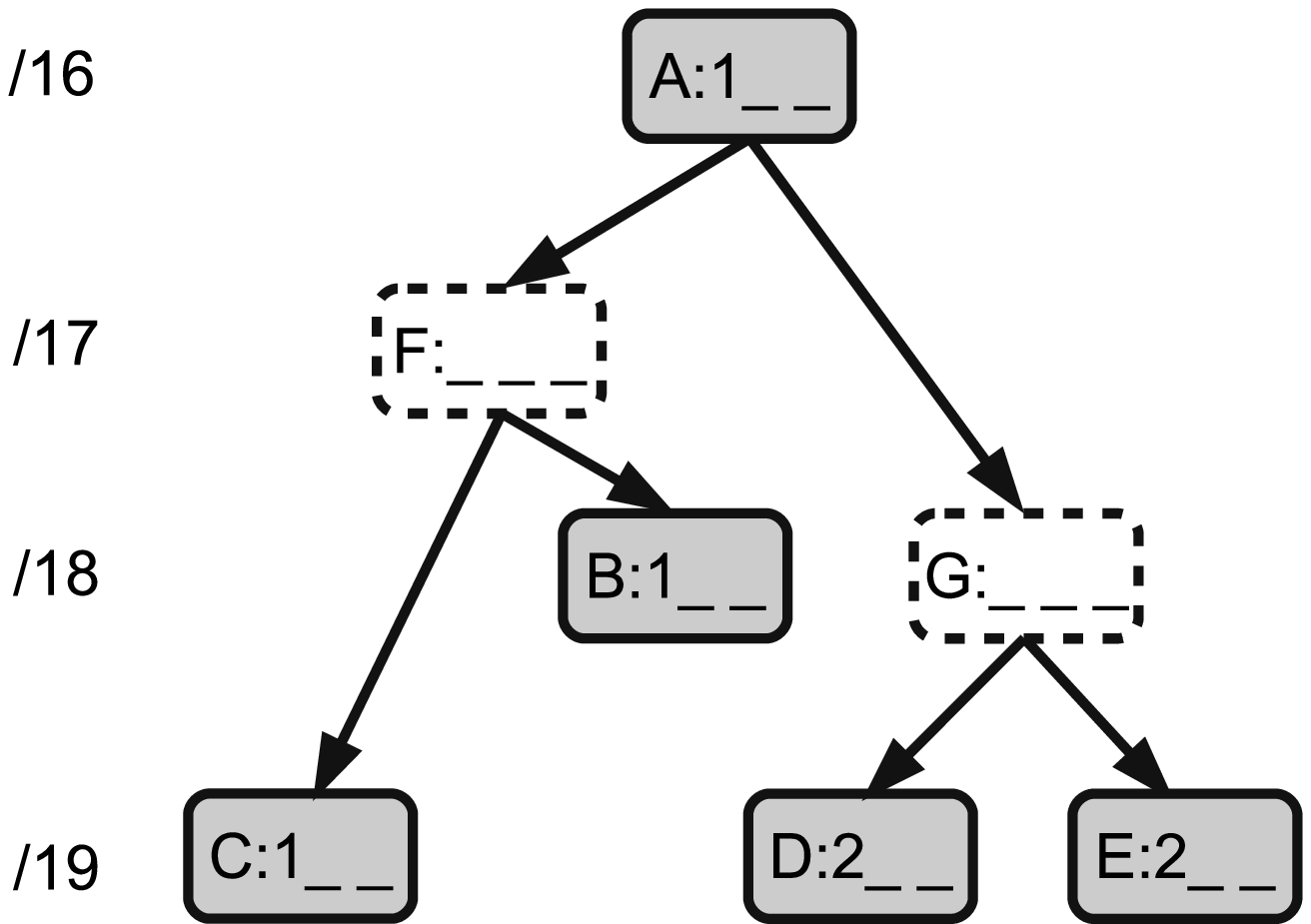}
	\label{fig:PatriciaI}
	\subfloat[PT after the top-down process]
	\centering
	\includegraphics[width=.25\linewidth]{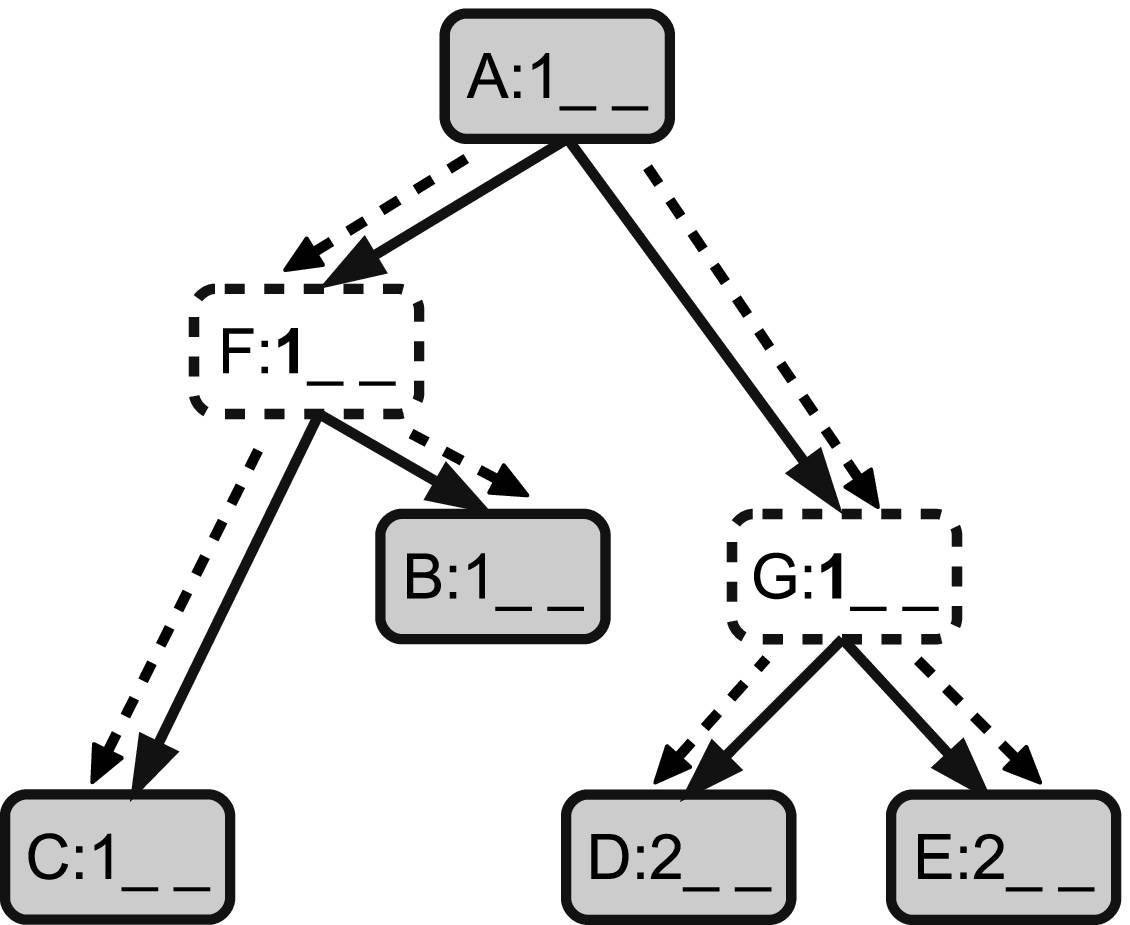}
	\label{fig:Patricia_top_down}
	\subfloat[PT after the bottom-up process]
	\centering
	\includegraphics[width=.25\linewidth]{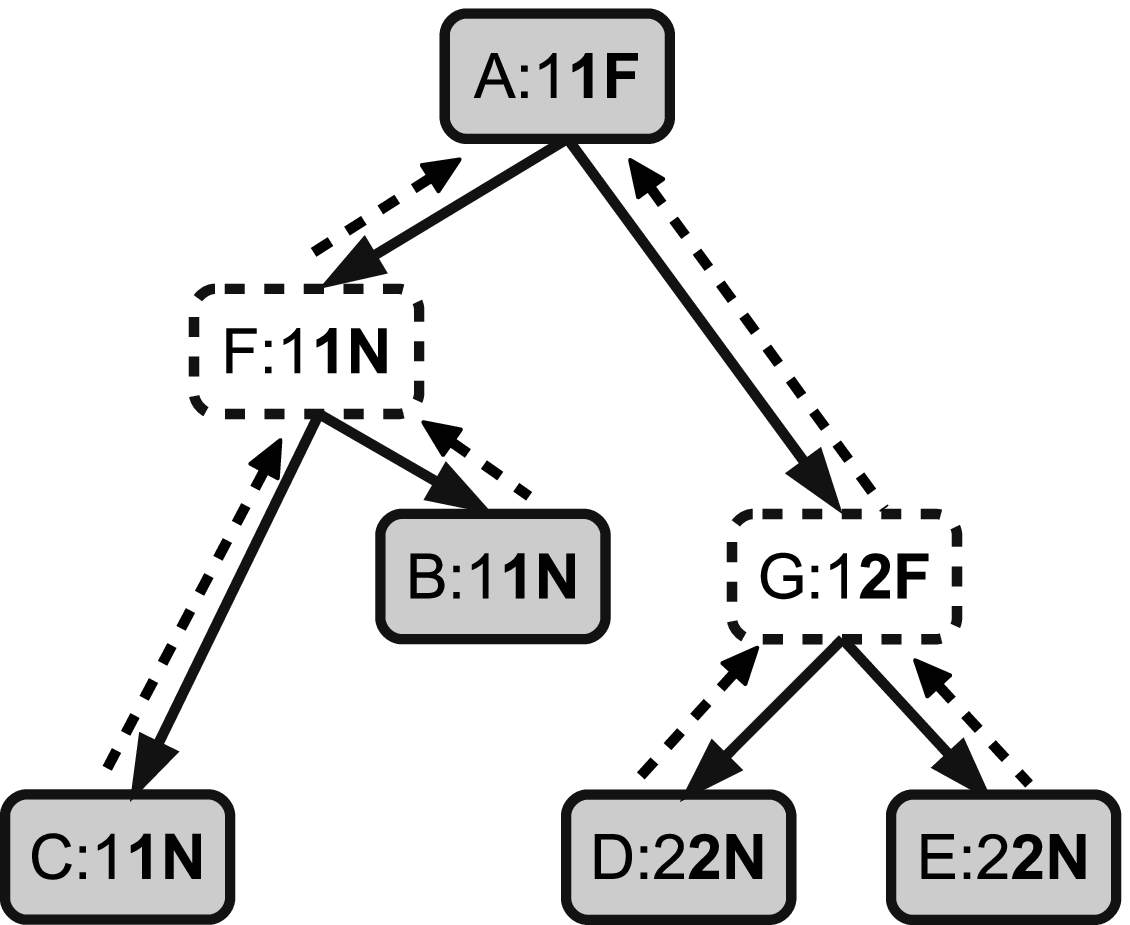}
	\label{fig:Patricia_bottom_up}
	\captionsetup{justification=centering, width=.9\linewidth}
	\caption{Static FIB aggregation of FIB from Table~\ref{tab:originalfibentry}. \\Fields in a node: (1) original next hop; (2) selected next hop, (3) FIB status: (F:IN\_FIB, N:NON\_FIB). \\ The solid nodes 
		denote \textit{REAL} nodes, whose prefixes are from the original FIB. 
	}
	\label{fig:Patricia}
	\vspace{-5mm}
\end{figure*}

\subsection{Static FIB aggregation}
\textit{FAQS} uses a data structure based on the PATRICIA trie (PT)~\cite{morrison1968patricia}. Each node in the PT has the following fields (we assume the current node labeled as $n$):

(1) \textit{Node type}, denoted by $T(n)$. If a node was derived from an original FIB entry, the value is \textit{REAL}; otherwise, if a PT node is only an ancillary node that helps to form the PT, the value is \textit{FAKE}. In Figure~\ref{fig:Patricia}(a), $T(F)$=$T(G)$=\textit{FAKE}, and $T(A)$=$T(B)$=$T(C)$=$T(D)$=$T(E)$=\textit{REAL}. 

(2) \textit{Original next hop}. The next hop value that is associated with an original FIB prefix and mapped to a PT node, denoted by \textit{O(n)}. For a \textit{REAL} node, it is taken from the FIB; for a \textit{FAKE} node, it is derived from the original hop of its nearest \textit{REAL} ancestor node during the top-down instantiation described below. 

(3) \textit{Selected next hop}. The next hop value of a prefix after aggregation, denoted by \textit{S(n)}. Note that a \textit{selected next hop} may be different from an \textit{original next hop} for the same prefix as long as aggregated FIB has exactly the same forwarding behaviors as the original one. 

(4) \textit{FIB status}, denoted by \textit{F(n)}. Indicates whether the prefix and its {selected next hop} should be placed in the FIB or not after FIB aggregation.  \textit{F(n)} can be equal to \textit{IN\_FIB} or \textit{NON\_FIB}. All routes with the status \textit{F(n)}=\textit{IN\_FIB} account for the entire aggregated FIB.  

After the initial PT is built from an original FIB, the nodes corresponding to its prefixes have the original next hop from FIB, the \textit{REAL} node type, and an empty selected next hop. The auxiliary nodes in the PT have an empty original next hop, the \textit{FAKE} type, and an empty selected next hop. Starting from here as shown in Figure~\ref{fig:Patricia}(a), \textit{Static FIB Aggregation} uses one-time post-order traversal to complete the whole aggregation, which consists of a recursive top-down and bottom-up stage.

$\bullet$ \textbf{Post-order top-down instantiation for an original next hop:}
For simplicity, the root node in the PT has the prefix 0/0 and the \textit{REAL} type. Its original next hop is either derived from the original FIB (if the FIB has a default next hop), or equal to 0 (to indicate a packet drop). From the root node of the PT, we instantiate the original next hop of each \textit{FAKE} node \textit{n} based on  \textit{O(n)=O(n.ancestor)}, where \textit{n.ancestor} is \textit{n}'s nearest \textit{REAL} ancestor. Figure~\ref{fig:Patricia}(b) shows the results after top-down process. The next hops of \textit{FAKE} nodes \textit{O(F)} and \textit{O(G)} are derived from the nearest \textit{REAL} ancestor $A$. 

$\bullet$ \textbf{Post-order bottom-up assignment for selected next hop and FIB status:} The bottom-up process consists of two operations for each node: assigning a node's selected next hop and determining the FIB status of its children. The selected next hop is assigned as follows:

I. Leaf nodes: \textit{S(n)=O(n)}.

II. Internal nodes: 

(1) \textit{S(n)=S(n.l)}, when the following conditions are satisfied: \textit{O(n)!=S(n.r)}, \textit{len(n.l)-len(n)=1} and \textit{len(n.r)-len(n)=1}, where \textit{n.l} and \textit{n.r} are node \textit{n}'s left and right child, and \textit{len(n)} represents the length of the prefix on node \textit{n}. Intuitively, the selected next hop value equals to its left child's selected next hop, when this node has two children nodes and the prefix length differences between this node and both of its children are exactly one, and the right child's selected next hop is different from its own original next hop. 


(2) \textit{S(n)=O(n)} in other cases. There can be three cases: (a) \textit{n} misses a child node; (b) The length of a child's prefix is longer than that of this node by more than 1; and (c) The selected next hop of a right child equals to the original next hop of \textit{n}. 


The next step for the bottom-up process is determining the FIB status of each node's children nodes. Assume \textit{n.l} and \textit{n.r} denote directly connected children of a node \textit{n}. Then,

(1) \textit{F(n.l)=IN\_FIB}, if \textit{n.l} exists and \textit{S(n.l)!=S(n)}.

(2) \textit{F(n.r)=IN\_FIB}, if \textit{n.r} exists and \textit{S(n.r)!=S(n)}.

Otherwise, children's node status will be \textit{NON\_FIB}.

Intuitively, we start aggregation from the leaf prefixes and recursively assign selected next hops based on their original next hops. When a child's selected next hop is the same as its parent's, the child's prefix and selected next hop can be excluded from the aggregated FIB. The process stops at the root node, which is always \textit{IN\_FIB}. The resultant aggregated FIB will have exactly the same forwarding behaviors as the original one. Figure~\ref{fig:Patricia}(c) shows the results after the bottom-up process. Algorithms~\ref{aggrround},~\ref{pselect}, and~\ref{pfib} present the pseudo code for the static FIB aggregation process. Finally, Table~\ref{tab:aggrfib1} illustrates the aggregated results, where the original five FIB entries are aggregated into two.

\begin{algorithm}
	\caption{Static FIB Aggregation}\label{aggrround}
	\begin{algorithmic}[1]
		\Procedure{$StaticAggregation$}{$node$}
		\State $p\gets node.parent$
		\State $l\gets node.left$
		\State $r\gets node.right$
		\If{$T(node)\not=REAL$}
		\State $O(node)\gets O(p)$
		\EndIf
		\If{$l\not=NULL$}
		\State $StaticAggregation(l)$ 
		\EndIf
		\If{$r\not=NULL$}
		\State $StaticAggregation(r)$ 
		\EndIf
		\State $setSelectedNexthop(node)$
		\State $setChildFIBstatus(node)$
		\EndProcedure
	\end{algorithmic}
\end{algorithm}

\begin{algorithm}
	\caption{Assignment of Selected Next Hop}\label{pselect}
	\begin{algorithmic}[1]
		\Procedure{$SetSelectedNexthop$}{$node$}
		\State $l\gets node.l$
		\State $r\gets node.r$
		\If{$l\not=NULL \land r\not=NULL \land $ \par $ len(l) - len(node)=1 \land $ \par $ len(r) - len(node) =1 \land  $ \par $ O(node)\not= S(r)$}
		\State $S(node)\gets S(l)$
		\Else
		\State $S(node)\gets O(node)$
		\EndIf
		\EndProcedure
	\end{algorithmic}
\end{algorithm}


\begin{algorithm}
	\begin{algorithmic}[1]
		\Procedure{$SetChildFIBstatus$}{$node$}
		\State $l\gets node.l$
		\State $r\gets node.r$
		\If{$l\not=NULL$}
		\If {$S(node)\not=  S(l)$}
		\State $F(l) \gets IN\_FIB$
		\Else \State $F(l) \gets NON\_FIB$
		\EndIf
		\EndIf
		\If{$r\not=NULL$}
		\If {$S(node)\not= S(r)$}
		\State $F(r) \gets IN\_FIB$
		\Else 
		\State $F(r) \gets NON\_FIB$
		\EndIf
		\EndIf
		\EndProcedure
	\end{algorithmic}
	\caption{Determine FIB status for Children Nodes}\label{pfib}
\end{algorithm}



\begin{table}[tbp]
	\centering
	\small
	\captionsetup{justification=centering}
	\caption{FIB entries after Aggregation by\\ FAQS}
	\label{tab:aggrfib}
	\begin{tabular}{| c | r | c | }
		\hline
		Label & Prefix  & Next Hop   \\ \hline
		A & 141.92.0.0/16 & 1    \\ \hline
		G & 141.92.192.0/18 & 2    \\ \hline
	\end{tabular}
	\label{tab:aggrfib1}
	\vspace{-3mm}
\end{table} 




\subsection{Incremental FIB update handling}
FIB updates consist of two categories: (a) Route announcements, including new routes and route changes, and (b) Route withdrawals. 

$\bullet$ \textbf{Route announcements: }
If the announced route is a new route, \textit{FAQS} algorithm generates a \textit{REAL} node with the corresponding original next hop in the PT; if it is a route update, it simply changes the original next hop value accordingly. In order to maintain a good aggregation ratio and forwarding correctness, the aggregated FIB needs to be re-aggregated. In FAQS, two portions of the PT may be affected: the subtree rooted at the updated node and the ancestors upon it. Specifically, the original next hop, the selected next hop and the FIB status of each node under the subtree need to be checked and updated if necessary. The process is similar to the procedure of the static FIB aggregation for the entire PT. Also, the selected next hop and the FIB status of each ancestor need to be checked and refreshed if necessary to maintain forwarding correctness. The procedure seems to be tedious, however, we leverage the following three crucial optimization techniques to greatly reduce the overall time costs and memory access times. 

(a) When adding a \textit{REAL} node or updating a \textit{FAKE} node, if the original next hop of this node's parent \textit{O(n.parent)} is same as the new next hop of the updated node \textit{O(n)}, then the top-down process can immediately terminate, since a parent's original next hop in the subtree rooted at \textit{n} does not change.

(b) Similarly, during the period of updating the subtree, if the node type of a node  \textit{T(n)} is \textit{REAL}, then the top-down process can stop on the current branch, because the original next hop of that node does not change. 

(c) During the period of updating the ancestors, if the newly selected next hop of an ancestor \textit{n} is the same as the old one before the update, then the bottom-up traversal can stop. Since update only happens on one branch and a parent's selected next hop is determined by its children's selected next hop, the preservation of a selected next hop of a node \textit{n} guarantees the invariance of all nodes above it.

\begin{figure*}[t!]
	\centering
	\captionsetup{justification=centering, width=.28\linewidth}
	\subfloat[Update node $D$ with the new next hop 3]
	\centering
	\label{fig:updatesS0}
	\includegraphics[width=.28\linewidth]{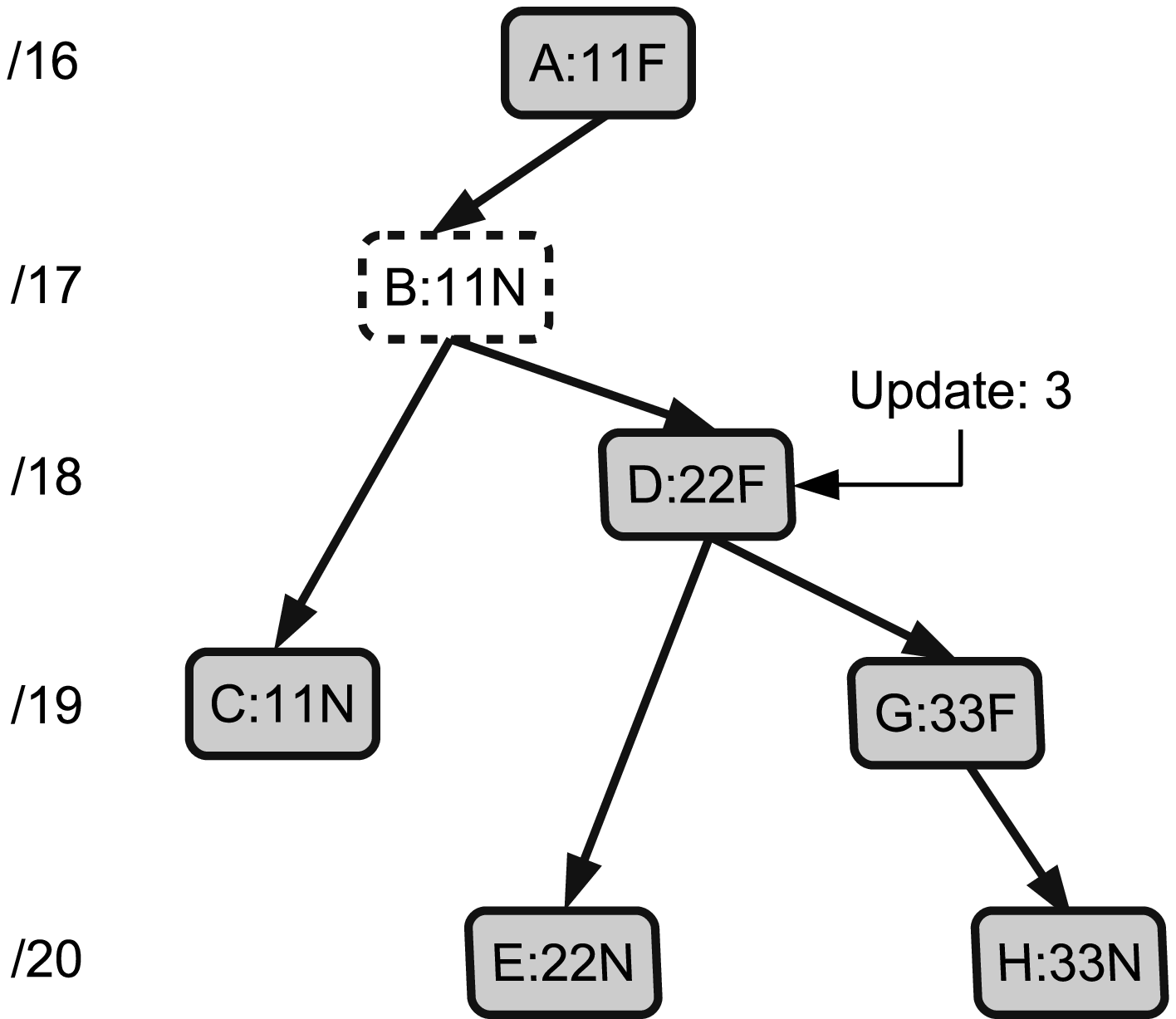}
	\captionsetup{justification=centering, width=.21\linewidth}
	\subfloat[Update original next hop to 3 and free selected next hop and FIB status]
	\centering
	\includegraphics[width=.23\linewidth]{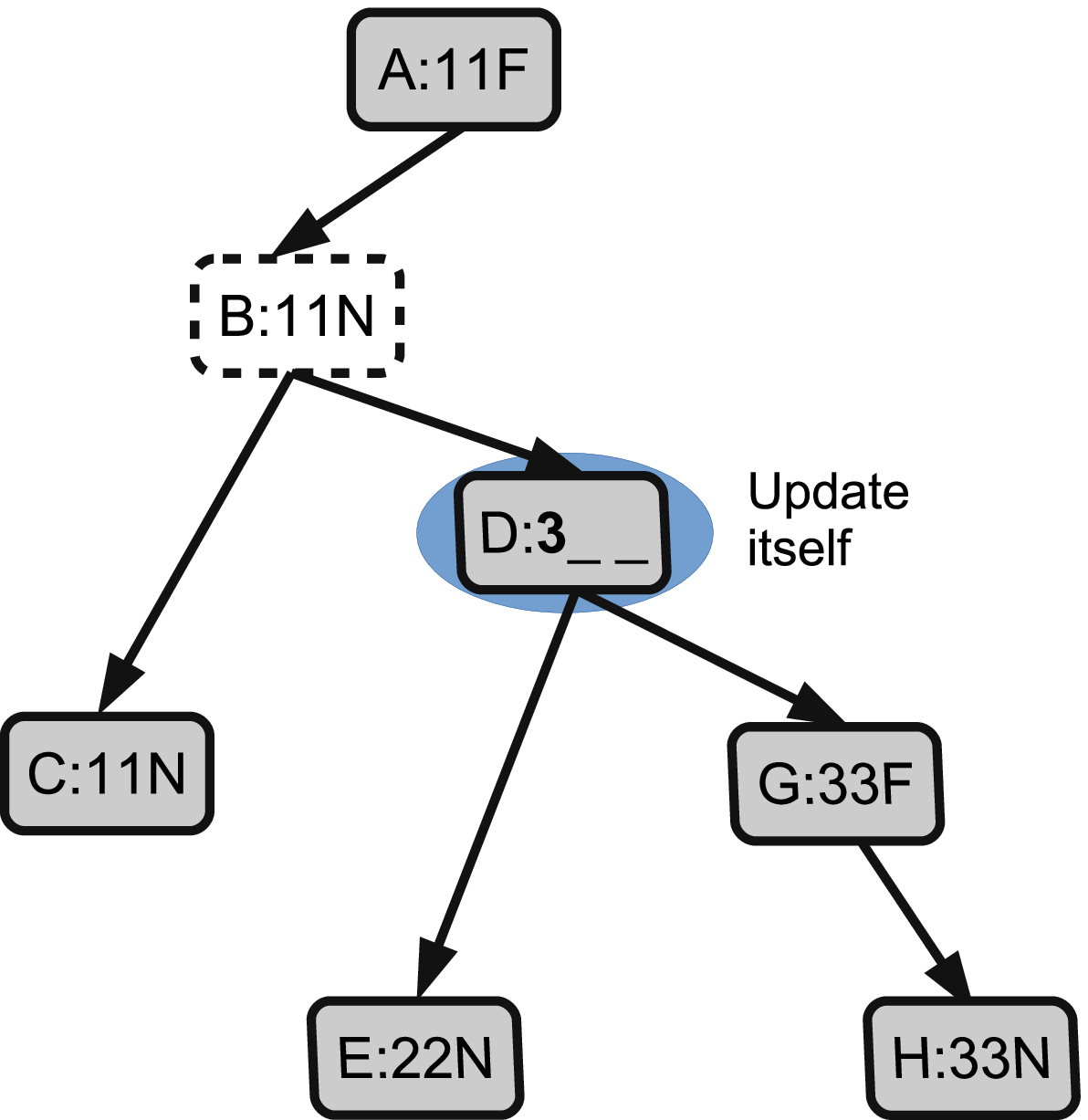}
	\label{fig:updatesS1}
	\subfloat[Update subtrie rooted at updated node $D$ and stop at \textit{REAL} node $G$]
	\centering
	\includegraphics[width=.23\linewidth]{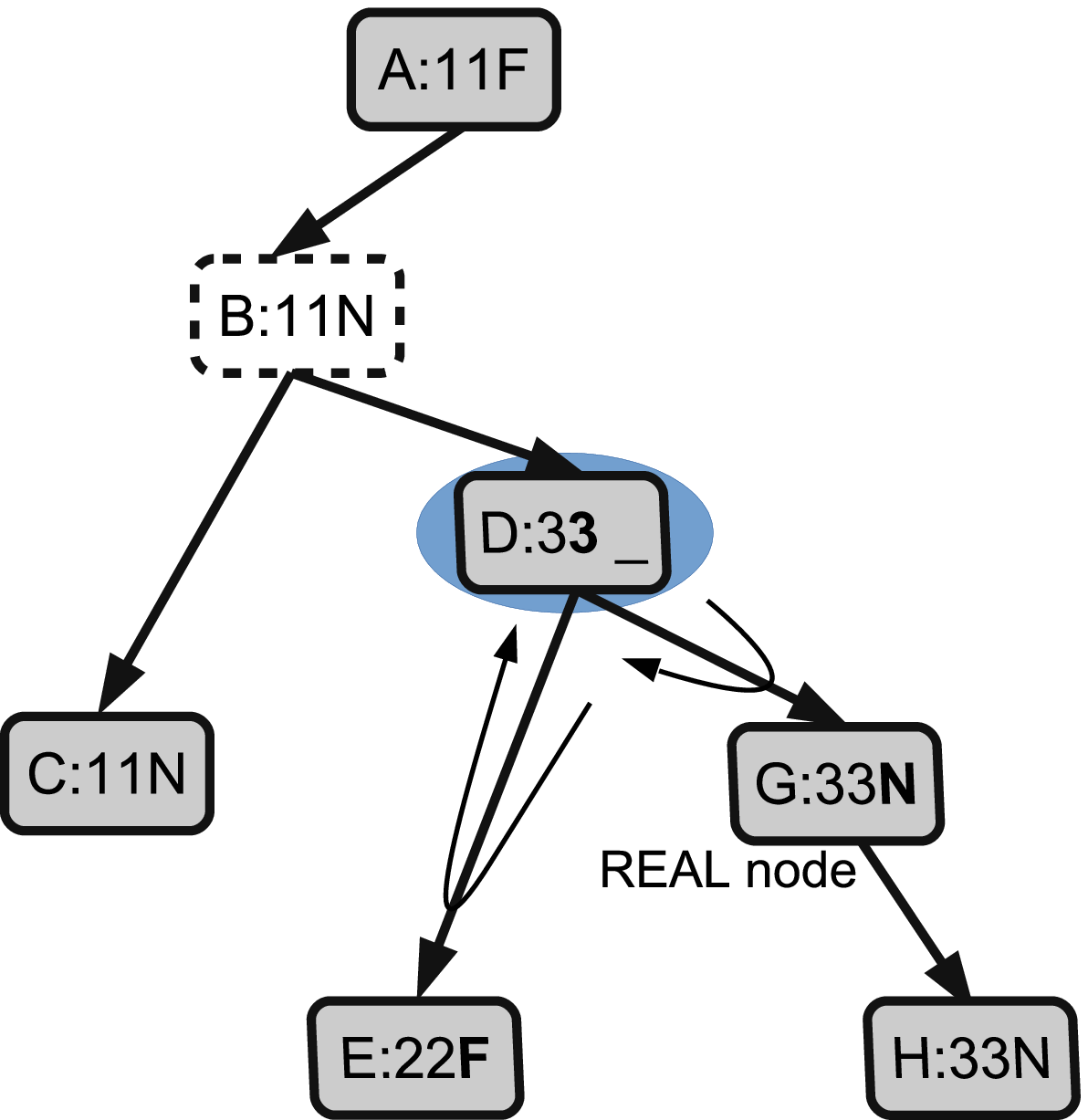}
	\label{fig:updatesS2}
	\subfloat[Update ancestors and stop when encountering the same selected next hop at node $B$]
	\centering
	\includegraphics[width=.23\linewidth]{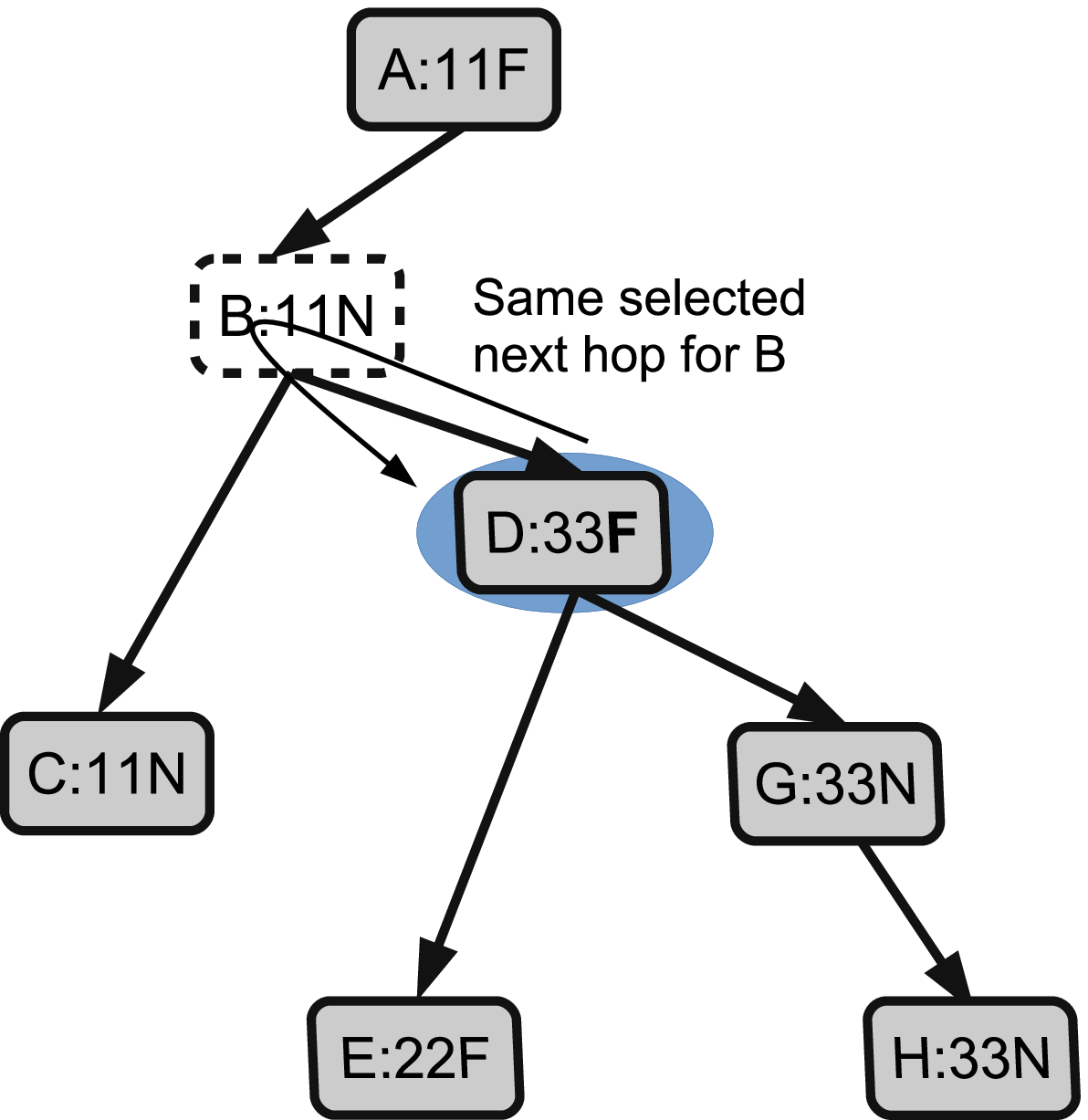}
	\label{fig:updateS3}

	\captionsetup{justification=centering, width=.9\linewidth}
	\caption{Incremental FIB update handling by FAQS}
	\label{fig:updatesS}
	\vspace{-4mm}
\end{figure*}

Algorithms~\ref{pa},~\ref{ppostorder} and~\ref{pb} illustrate the whole process of incremental update handling and Figure~\ref{fig:updatesS} demonstrates an example to update a route with a new next hop, where the second and third optimization techniques are applied. In the example, Node $D$ has an update with a new next hop 3. First the original next hop changes to 3 and other fields are freed; then the update-tree process stops when encountering a \textit{REAL} node $G$. After that, the update-ancestor process stops when the same selected next hop 1 is discovered at node $B$. As a result, we can observe that only a small portion of the trie has been traversed to incrementally handle the update.

\begin{algorithm}
	\caption{Incremental FIB Update Handling}\label{pa}
	\begin{algorithmic}[1]
		\Procedure{$UpdateNode$}{$prefix,nexthop$}
		\State $node\gets find(prefix)$
		\If{$node=NULL$}
			\State $node\gets initialize(prefix,nexthop)$
			\State $T(node)\gets REAL$
			\State $p\gets node.parent$
		\If{$O(p)\not=O(node)$}
			\State $UpdateSubtree(node)$
			\State $UpdateAncestors(node)$
		\EndIf
		\Else
		\If {$T(node) \not= REAL$}
			\State $T(node)\gets REAL$
		\EndIf
		\If {$ O(node) \not= nexthop$}	
			\State $O(node)\gets nexthop$
			\State $UpdateSubtree(node)$
			\State $UpdateAncestors(node)$
		
		\EndIf
		\EndIf
		\EndProcedure
	\end{algorithmic}
\end{algorithm}

\begin{algorithm}
	\caption{Update Subtree}\label{ppostorder}
	\begin{algorithmic}[1]
		\Procedure{$UpdateSubtree$}{$node$}
		\State $l\gets node.l$
		\State $r\gets node.r$
		\If{$l\not=NULL \land T(l)\not=REAL$}
		\State $O(l)\gets O(node)$
		\State $UpdateSubtree(l)$ 
		\EndIf
		\If{$r\not=NULL \land T(r)\not=REAL$}
		\State $O(r)\gets O(node)$
		\State $UpdateSubtree(r)$ 
		\EndIf
		\State $setSelectedNexthop(node)$
		\State $setChildFIBstatus(node)$
		\EndProcedure
	\end{algorithmic}
\end{algorithm}

\begin{algorithm}
	\begin{algorithmic}[1]
		\Procedure{$UpdateAncestors$}{$node$}
		\State $p \gets node.parent$
		\While{$p \not= NULL$}
		\State $oldSlctNexthop \gets S(p)$
		\State $setSelectedNexthop(p)$
		\State $setChildFIBstatus(p)$
		\If{$oldSlctNexthop = S(p)$}
		\State $break$
		\EndIf
		\State $p \gets p.parent$
		\EndWhile
		\EndProcedure
	\end{algorithmic}
	\caption{Update Ancestors}\label{pb}
\end{algorithm}

$\bullet$ \textbf{Route withdrawal: }
The \textit{FAQS} algorithm handles the prefix withdrawals within two steps:

(a) Node removal. First, \textit{FAQS} looks up the corresponding \textit{REAL} node from the PT. If the node is found, then \textit{FAQS} checks if it is removable. A removable node refers to a node, which will not affect the PT structure after its deletion. In such case, \textit{FAQS} deletes the node and reorganizes the pointers of its parent and child. Otherwise, if the node is not removable, \textit{FAQS} changes its type to \textit{FAKE} and frees the values of the original next hop, the selected next hop and the FIB status.

(b) Trie update. Starting from the parent node of the deleted or updated node, the incremental update process will be the same as the case of route announcements. First, \textit{FAQS} does a top-down update of the original next hops of nodes on the subtree; next, it bottom-up updates the values of the selected next hops and the FIB status for each node all the way to the point where a new selected next hop does not change. The three optimization techniques used in route announcements apply here as well.

%% file: results.tex
\section{Evaluation}
\label{sec:evaluation}

We used realistic IPv4 and IPv6 routing tables from 2011 to 2016 in Route Views project~\cite{routeviews} for the evaluation. We collected one baseline routing table on 01/01/2011 for both IPv4 and IPv6, and applied all following updates to obtain the aggregation results. We use AS neighbors as the next hops for FIB tables, because local FIB interface information is not available in the dataset. Normally, the number of interfaces in a FIB is much less than the number of its neighbors. Thus our results underestimate the real FIB aggregation effects. We verified the forwarding behaviors before and after aggregation and they are equivalent. We ran our experiment on an Intel Xeon Processor E5-2603 v3 1.60GHz machine. 
We compared our \textit{FAQS} algorithm with the optimal ORTC-based \textit{FIFA-S}~\cite{liu2013fifa} aggregation algorithm. Unlike~\textit{FIFA-T}, a faster version of \textit{FIFA} algorithms, \textit{FIFA-S} has significantly smaller FIB bursts, which is critical since writing operations on TCAM are slow~\cite{bifulco2015towards}.

We used the following metrics for our experiment: 

\begin{enumerate}
	
	\item \textit{FIB Size}: the total number of entries before and after aggregation. \textit{Aggregation Ratio} is calculated by the ratio between the total number of the FIB entries after aggregation and before aggregation.
	\item \textit{FIB Aggregation Time}: the time spent handling all route updates by the aggregation algorithm (before pushing FIB changes into the data plane).
	\item \textit{Total Number of FIB Changes}: the total number of FIB changes that are pushed into the data plane by the aggregation module upon handling all route updates. One route update from the control plane may result in zero or more FIB changes to the data plane FIB due to the incremental FIB aggregation process. If there is no aggregation, one route update corresponds to one FIB change. 
	\item \textit{FIB Burst}: The number of FIB changes caused by a single route update, either a route announcement or a withdrawal. 

\end{enumerate}

\subsection{IPv4 results}
We use five routing tables from different ASes to demonstrate the dependency of aggregation performance on the number of neighbors (i.e. the number of possible next hops). The number of next hops ranges from 21 to 4500. To illustrate the worst case, we use a routing table in AS3356 that has 4500 next hops on 12/31/2016. There are more than 426 million route updates to be handled for the 6-year period. 

\begin{figure*}[t]
	\begin{multicols}{3}
		\centering
		\captionsetup{justification=centering, width=\linewidth}
		\subfloat[FIB Size]
		\centering
		\includegraphics[width=\linewidth]{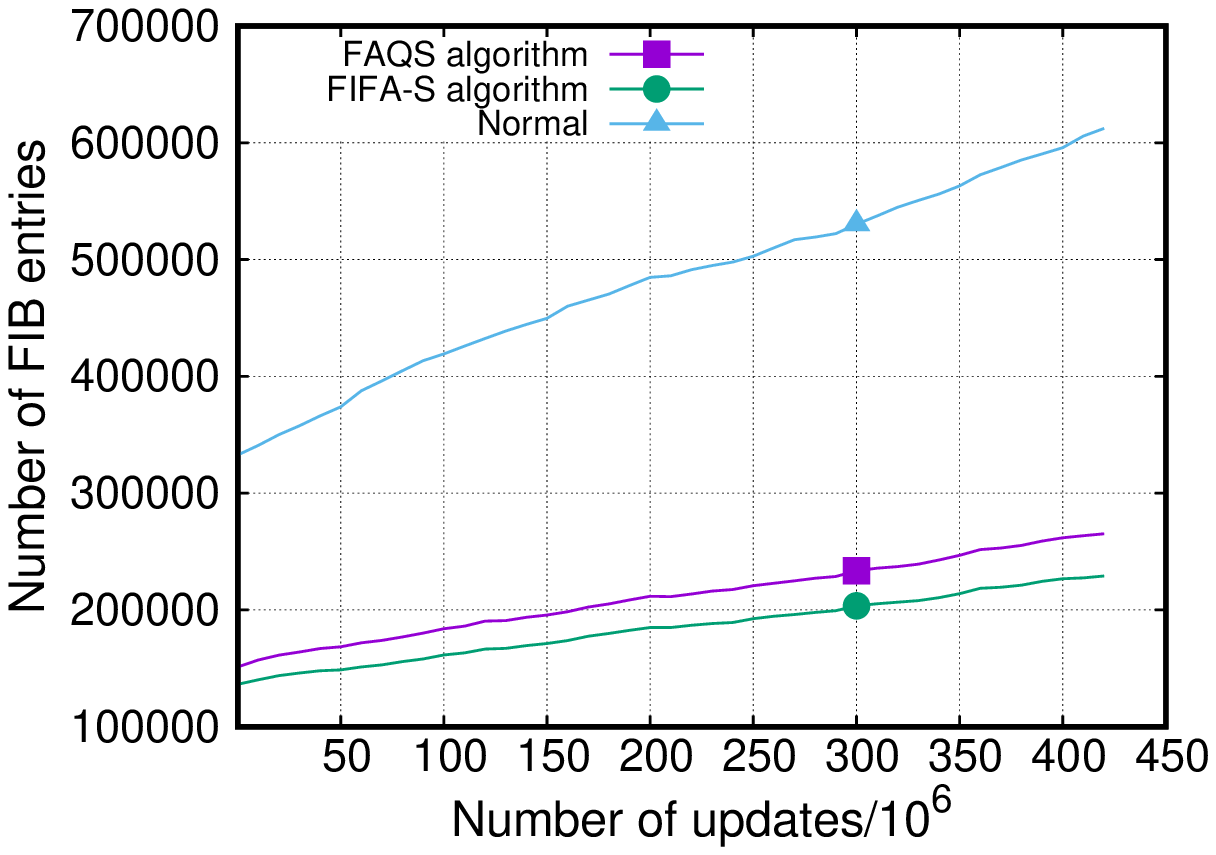}
		\label{fig:ipv4wfibsize}
		\subfloat[FIB Aggregation Time]
		\centering
		\includegraphics[width=\linewidth]{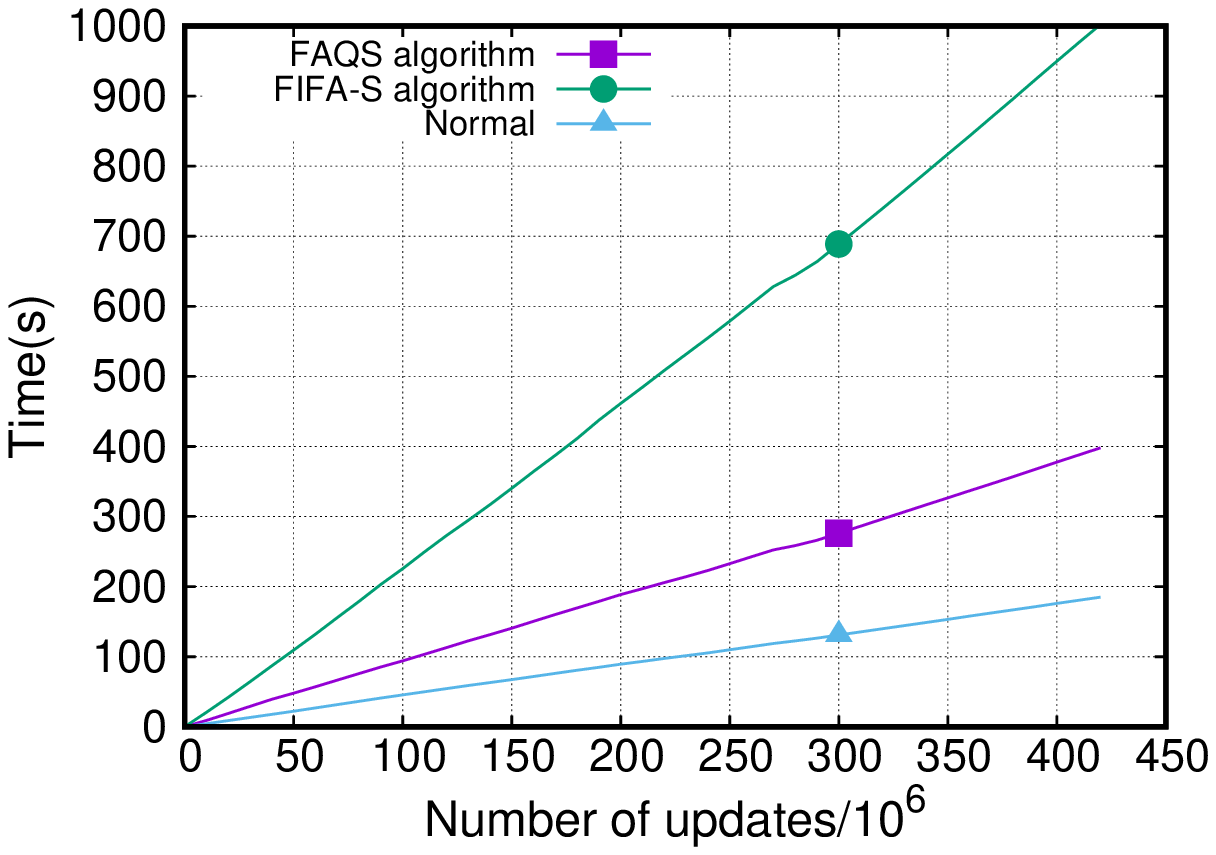}
		\label{fig:ipv4fibtime}	
		
		\subfloat[FIB Changes]
		\centering
		\includegraphics[width=\linewidth]{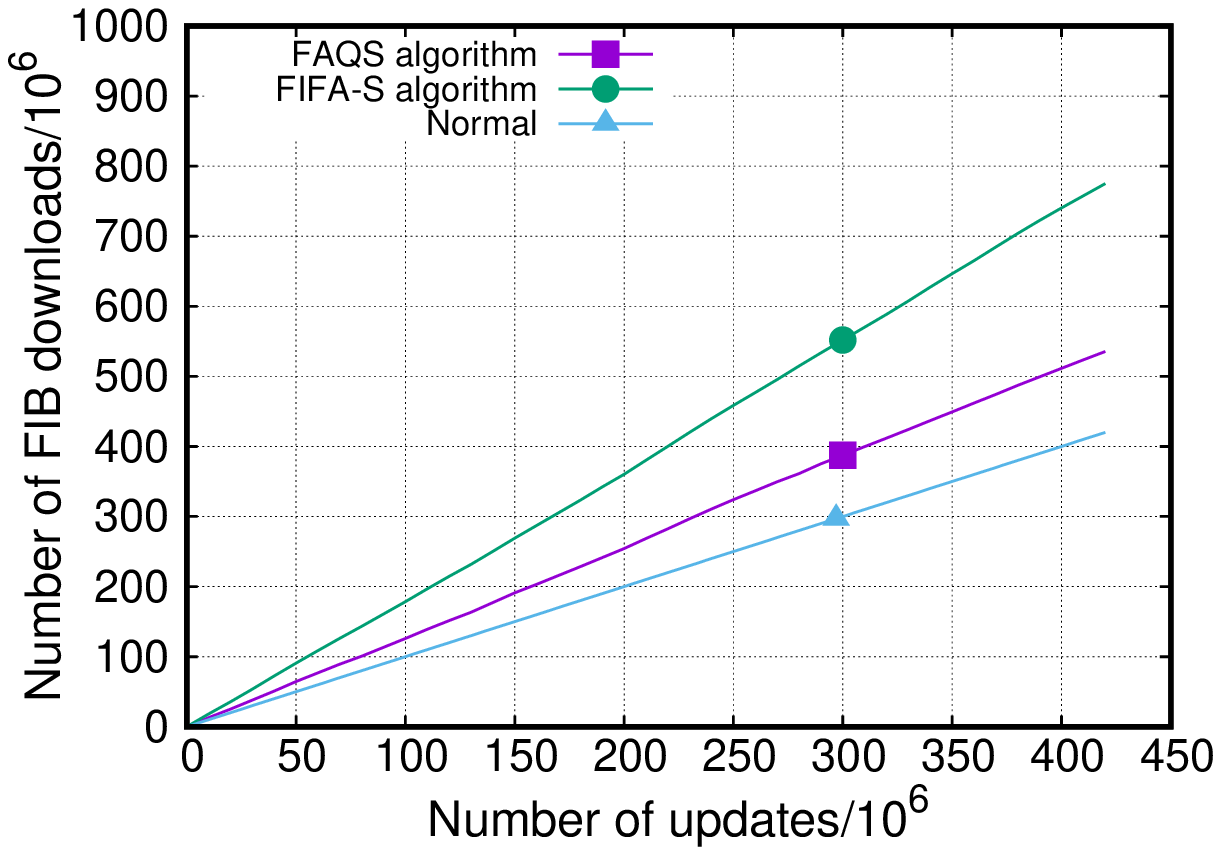}
		\label{fig:ipv4wfibdownloads}
		
	\end{multicols}
	\vspace{-5mm}	
	\captionsetup{justification=centering, width=\linewidth}
	\caption{FIB aggregation of IPv4 routing table (AS 3356)}
	\label{fig:ipv4worstcase}
	\vspace{-4mm}
\end{figure*}

Figure~\ref{fig:ipv4worstcase}(a) shows the number of FIB entries without aggregation, using \textit{FIFA-S} algorithm and \textit{FAQS} aggregation algorithms.  The top green line marked by a triangle represents the FIB size without aggregation. The middle line marked by a rectangle represents the FIB size after \textit{FAQS} and the bottom line represents the FIB size after \textit{FIFA-S}. Both of the aggregation algorithms can compress the original FIB by around 60\%. Since \textit{FIFA-S} reaches optimal aggregation ratio for each route update, \textbf{\textit{FAQS} can achieve near-optimal aggregation ratio}. 

However, \textit{FAQS} uses much less time to complete the aggregation as shown in Figure~\ref{fig:ipv4worstcase}(b). \textit{FIFA-S} takes around 1000s to finish with an average 2.38$\mu$$s$ per update, while \textit{FAQS} takes about 400s to finish with an average 0.94$\mu$$s$ per update. Thus \textbf{\textit{FAQS} is 2.53 times faster than \textit{FIFA-S} but bears similar aggregation ratio}. The primary reason is that \textit{FIFA-S} needs to traverse a subtree twice to handle an update with additional memory consumption but \textit{FAQS} only needs one-time traversal as described in Section~\ref{sec:design}. The numbers also indicate that \textit{FAQS} can handle more than 1 million updates per second and can be well adopted by Internet backbone routers, given that BGP churn can be up to 500,000 per minute~\cite{elmokashfi2012bgp}. 

The smaller number of FIB changes to the FIB, the better performance. Figure~\ref{fig:ipv4worstcase}(c) shows that \textbf{\textit{FAQS} algorithm generates 31\% less number of FIB changes than that of \textit{FIFA-S} algorithm} (543,309,259 vs 786,633,132). The average number of FIB changes per update is 1.27 for \textit{FAQS} and 1.84 for \textit{FIFA-S}. Both algorithms have similar distribution for the size of FIB bursts as shown in Table~\ref{tab:Summary}(a). The vast majority of FIB bursts (more than 99.97\%) in both algorithms consist of 30 FIB changes and less. The largest FIB burst for \textit{FAQS} is 1443, which is slightly smaller that \textit{FIFA-S} (1496). Nonetheless, \textbf{the update handling time cost for the largest burst in \textit{FAQS} takes only 30\% of running time of \textit{FIFA-S}}. Table~\ref{tab:Summary}(a) presents other evaluation results of FIB aggregation for the five ASes. It is interesting to observe that a good percentage (6.05\%-14.91\%) of FIB updates result in zero FIB changes (column $n_b$=0). 

%
%
%

%

\begin{figure*}[t]
	\begin{multicols}{3}
		\centering
		\captionsetup{justification=centering, width=\linewidth}
		\subfloat[\label{fig:ipv6wfibsize}FIB Size]
		\centering
		\includegraphics[width=\linewidth]{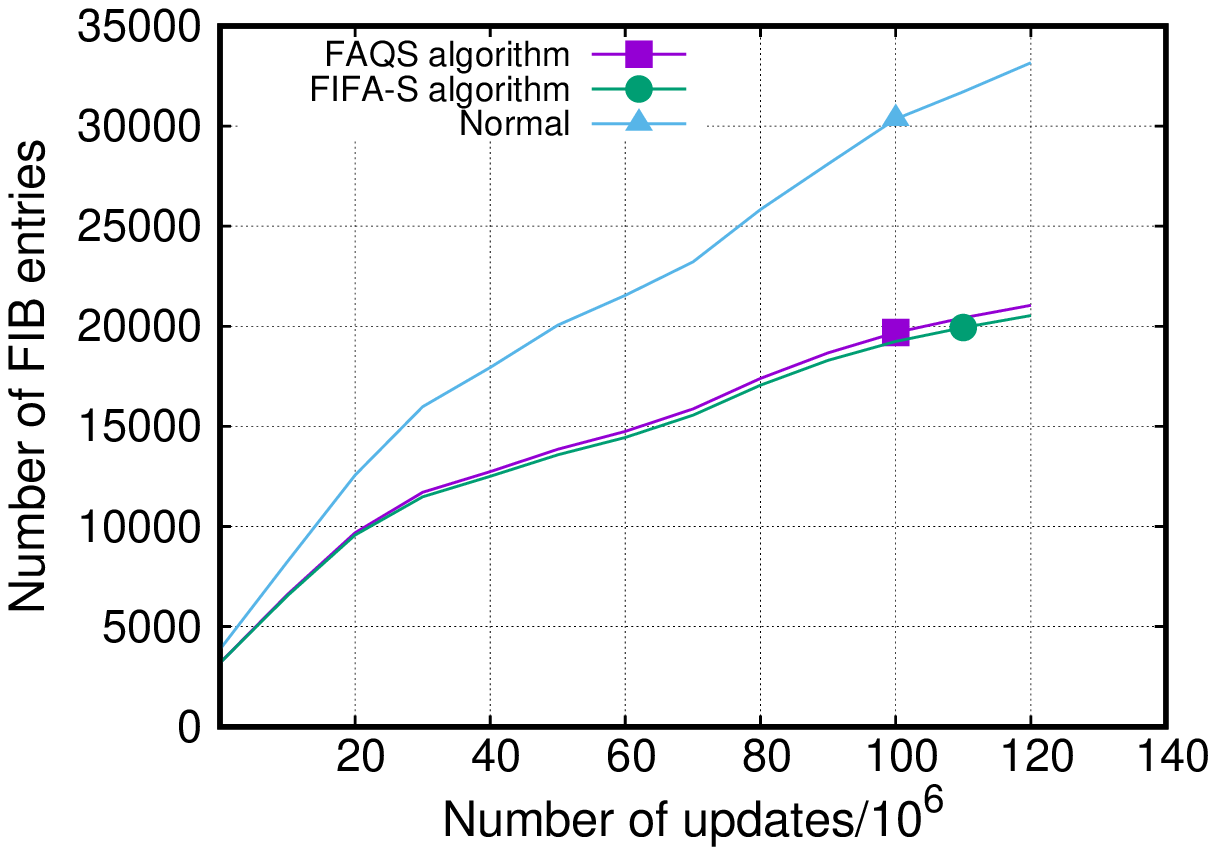}
		\subfloat[\label{fig:ipv6fibtime}FIB Aggregation Time]
		\centering
		\includegraphics[width=\linewidth]{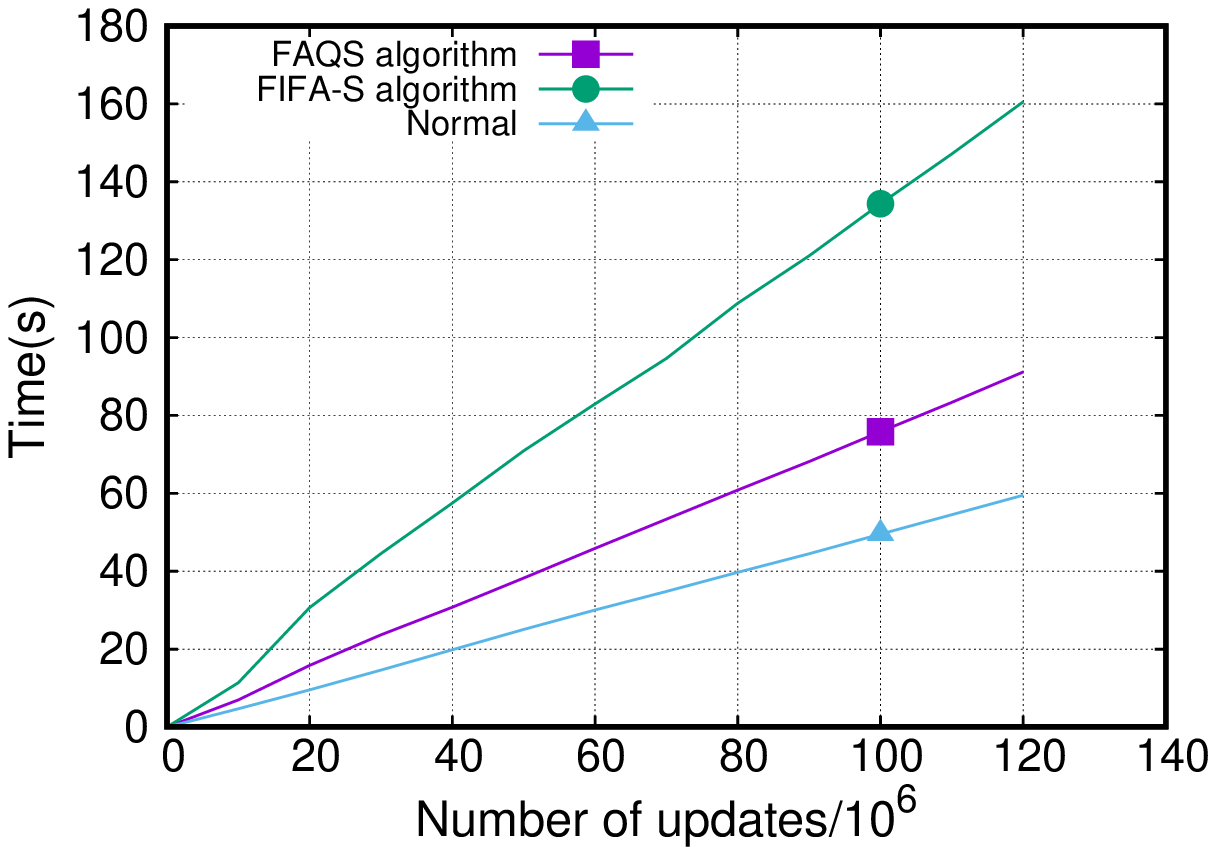}
		\subfloat[FIB Changes]
		\centering
		\includegraphics[width=\linewidth]{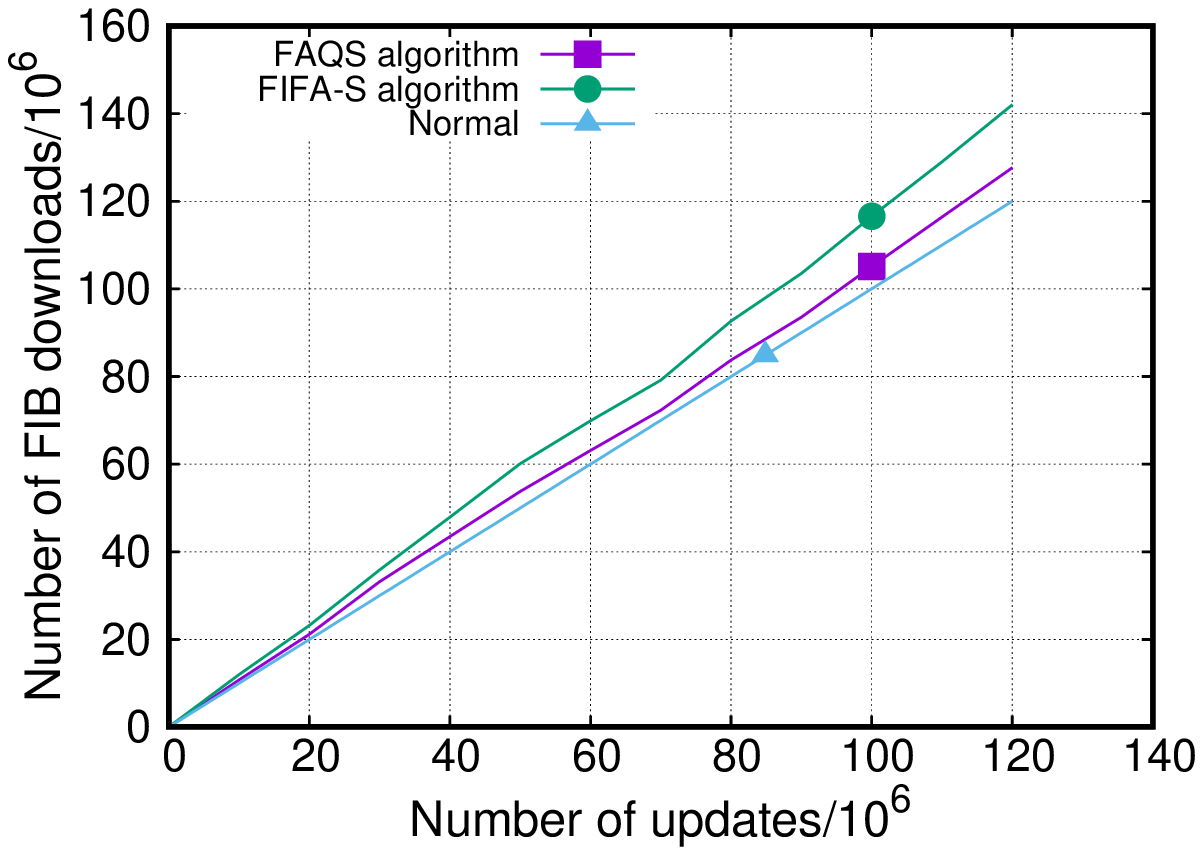}
		\label{fig:ipv6wfibdownloads}
	\end{multicols}
	\vspace{-6mm}	
	\captionsetup{justification=centering, width=\linewidth}
	\caption{FIB aggregation of IPv6 routing table (AS 6939)}
	\label{fig:ipv6worstcase}
	\vspace{-3mm}
\end{figure*}



\begin{table*}[tbp]
	\centering
	\subfloat[IPv4 routing tables]{
	\begin{tabular}{|r|r|r|c|c|r|r|r|r|r|r|r|c|}
		\hline
		AS & $n_{u}$ & $p_{avg}$ & Algorithms & $r$ & $n_{c}/n_{u}$ & $t_{aggr}$ & $t_{peak}$ &  $n_{b}=0$ & $n_{b}=1$ & $n_{b}\leq30 $ & $b_{max}$  & $m$  \\
		\hline
		\multirow{2}{*}{3356} & \multirow{2}{*}{426551755} & \multirow{2}{*}{3746} & FAQS & 0.43 & 1.27 & 0.94$\mu$s & 3.09$ms$ & 12.75\% & 56.21\% & 99.97\% & 1443 & 175MB \\
		\hhline{|~|~|~|----------} 
		& & & FIFA & 0.37 & 1.84 & 2.38$\mu$s & 10.29$ms$ & 12.85\% & 64.23\% & 99.98\% & 1496 & 228MB\\
		\hline
		\multirow{2}{*}{7018} & \multirow{2}{*}{782293331}  & \multirow{2}{*}{2397} & FAQS & 0.41 & 1.21 & 0.94$\mu$s & 3.09$ms$ & 16.43\% & 61.71\% & 99.99\% & 1854 & 175MB \\
		\hhline{|~|~|~|----------} 
		& & & FIFA & 0.34 & 1.78 & 2.06$\mu$s & 8.39$ms$ & 16.38\% & 54.57\% & 99.97\% & 1892 & 229MB \\
		\hline
		\multirow{2}{*}{8492} & \multirow{2}{*}{1037150247}  & \multirow{2}{*}{1126} & FAQS & 0.39 & 1.37 & 0.93$\mu$s & 3.27$ms$ & 6.05\% & 69.45\% & 99.97\% & 4268& 178MB \\
		\hhline{|~|~|~|----------} 
		& & & FIFA & 0.32 & 1.90 & 2.33$\mu$s & 12.14$ms$ & 6.40\% & 63.77\% & 99.97\% & 4657 & 233MB\\
		\hline
		\multirow{2}{*}{1239} & \multirow{2}{*}{295214072}  &\multirow{2}{*}{739} & FAQS & 0.42 & 1.28 & 1.09$\mu$s & 3.56$ms$ & 13.18\% & 62.18\% & 99.98\% & 1585 & 175MB\\
		\hhline{|~|~|~|----------} 
		& &  & FIFA & 0.36 & 1.91 & 2.69$\mu$s & 12.68$ms$ & 13.38\% & 55.03\% & 99.97\% & 1952 & 229MB\\
		\hline
		\multirow{2}{*}{3130} & \multirow{2}{*}{402445005}  &\multirow{2}{*}{23} & FAQS & 0.27 & 1.23 & 0.95$\mu$s & 3.26$ms$ & 14.69\% & 63.50\% & 99.98\% & 6464 & 174MB \\
		\hhline{|~|~|~|----------} 
		&&  & FIFA & 0.20 & 1.68 & 2.04$\mu$s & 16.02$ms$ & 14.91\% & 56.74\% & 99.98\% & 5524 & 228MB\\
		\hline
	\end{tabular}
		}

	\subfloat[IPv6 routing tables]{
		\begin{tabular}{|r|r|r|c|c|r|r|r|r|r|r|r|c|}
			\hline
			AS & $n_{u}$  & $p_{avg}$  & Algorithms & $r$ & $n_{c}/n_{u}$ & $t_{aggr}$ & $t_{peak}$ &  $n_{b}=0$ & $n_{b}=1$ & $n_{b}\leq30 $ & $b_{max}$ & $m$\\
			\hline
			\multirow{2}{*}{6939} & \multirow{2}{*}{122903741}  &\multirow{2}{*}{2725} & FAQS & 0.63 & 1.06 & 0.76$\mu$s & 1.27$ms$ & 7.08\% & 84.19\% & 99.99\% & 181 & 11MB\\
			\hhline{|~|~|~|----------} 
			&  & & FIFA & 0.61 & 1.18 & 1.33$\mu$s & 2.97$ms$ & 7.09\% & 81.19\% & 99.98\% & 258 & 14MB\\
			\hline
			\multirow{2}{*}{33437} & \multirow{2}{*}{33486605}  &\multirow{2}{*}{7} & FAQS & 0.58 & 0.98 & 0.90$\mu$s & 1.42$ms$ & 17.47\% & 73.68\% & 99.99\% & 2447 & 11MB \\
			\hhline{|~|~|~|----------} 
			&  & & FIFA & 0.56 & 1.11 & 1.43$\mu$s & 2.48$ms$ & 17.46\% & 68.33\% & 99.99\% & 2432 & 14MB \\
			\hline
		\end{tabular}
	}
	\captionsetup{justification=centering, width=\linewidth}
	\vspace{-2mm}
	\caption{Evaluation summary (2011-2016 period). $n_{u}$ - the number of FIB updates; $p_{avg}$ - average peer number;  $r$ - aggregation ratio; $n_{c}/n_{u}$ - the ratio between the number of FIB changes and FIB updates; $t_{aggr}$ - average aggregation time per update; $t_{peak}$ - peak aggregation time; $n_{b}$ - percentage of updates with burst values 0, 1 and below 30; $b_{max}$ - maximum burst value; $m$ - memory consumption.}
	\label{tab:Summary}
	\vspace{-6mm}
\end{table*}

\subsection{IPv6 results}

To the best of our knowledge, this is the first time that IPv6 routing tables have been evaluated for their aggregation results. We aggregated FIB tables from AS 6939 with 3501 next hops. The total number of route updates to be handled is more than 122 million. Figure~\ref{fig:ipv6worstcase} shows the curves of FIB size, aggregation time and the total number of FIB changes. In Figure~\ref{fig:ipv6wfibsize}, we can observe that the size of IPv6 routing tables has increased dramatically since six years back, when there were only less than 5,000 entries. In the end of 2016, it has been close to 35,000. Due to the small size, the aggregation ratios for both \textit{FAQS} and \textit{FIFA-S} are around 60\%, which are not as good as IPv4. Since \textit{FIFA-S} outputs the smallest aggregated FIB, \textit{FAQS}'s aggregation ratio for IPv6 is close to optimal. Remarkably, the running time of \textit{FAQS} is much lower than \textit{FIFA-S} (90s vs 160s in Figure~\ref{fig:ipv6fibtime}) while they have similar aggregation ratios, which again attributes to the one-time subtree traversal with three important optimization techniques for \textit{FAQS} while \textit{FIFA-S} uses two traversals.  Table~\ref{tab:Summary}(b) demonstrates results for both AS6939 and AS33437. AS33437 has only 7 next hops, thus the aggregation ratio is better (58\% vs 56\% for \textit{FAQS} and \textit{FIFA-S}, respectively) and the burst size is larger than the one in AS6939, because one update in AS33437 may affect a larger area of next hops.

%% file: related.tex
\section{Related Work}
\label{sec:related}
\vspace{-2mm}
A number of FIB aggregation algorithms have been proposed. We highlight a few of them here.
\textit{SMALTA} algorithm~\cite{Uzmi11:smalta} uses the binary tree data structure and bases on \textit{ORTC} algorithm~\cite{draves1999constructing}, which can achieve one-time optimal aggregation.
\textit{SMALTA} takes \textit{ORTC} as the initial FIB aggregation algorithm and processes updates without the optimization of a subtree, rooted at the updated node. Eventually, \textit{SMALTA} requires full re-aggregation of the FIB table upon reaching FIB size threshold. It results in computational spikes and high time costs. In~\cite{sarrar2014leveraging}, authors study and employ the locality of FIB updates to build Locality-aware FIB Aggregation (LFA) algorithm. In LFA, reaggregation for an updated prefix region is delayed until it is stabilized. However, such approach requires timers attached to nodes which may significantly complicate its operation in the real routers. Bienkowski et al.~\cite{bienkowski2014competitive} present a formal study on the trade-off between FIB aggregation and update bursts. In addition, paper presents the algorithm \textit{HIMS} that attaches time-dependent counters to each node as well. However, the paper provides no information on the performance of the algorithm when processing real network routing data. In~\cite{karpilovsky2012practical}, authors propose MMS, the Memory Management System designed to prolong the lifetime of legacy routers in an ISP. MMS uses parallelization of ORTC and can aggregate routing tables locally or on an AS-level. Moreover, MMS may change the forwarding behavior of routers in order to gain additional compression.

Some FIB compression work uses smart data structures to minimize storage size of FIB~\cite{retvari2012compressing}. In~\cite{retvari2013compressing}, authors present a tunable aggregation algorithm with compressed prefix trees. By changing the deepness of the compression, network operators can manage the trade-off between the aggregation ratio and BGP update overhead.  Similarly, Yang et al. in~\cite{Yang:2012:AOC:2330748.2330780} present two algorithms, \textit{EAP-slow} and \textit{EAP-fast} and compare it with \textit{ORTC}. In~\cite{luo2015practical}, authors propose an aggregation algorithm for OpenFlow flow tables using prefix wildcards. FIB aggregation scheme, that applies multiple selectable next hops, is  proposed by Li et al.~\cite{li2015nexthop}.  Abraham et al.~\cite{abraham2014flexible} create a virtual network system to implement and study FIB aggregation. It is a reusable framework to test the performance of FIB aggregation algorithms in a realistic environment. 

Aggregation algorithms such as \textit{Level-1} and \textit{Level-2}~\cite{Xin10:FIBAGGREGATION} compress FIB quickly but bear costly update handling operations.  In 2013, Liu et al. developed \textit{FIFA} algorithms~\cite{liu2013fifa}, which improves \textit{ORTC} algorithm by applying PATRICIA trie (PT) with incremental FIB aggregation features. 





Our work, \textit{FAQS} algorithm, makes a good balance of aggregation time, ratio and memory consumption. It sacrifices very little aggregation ratio compared with the optimal solution, but speeds up the aggregation more than twice with much less memory consumption. Considering the real-time and efficiency requirements of FIB aggregation, our approach is superior to the existing algorithms. 


%% file: conclusion.tex
\section{Conclusion}
\label{sec:conclusion}
\textit{FIB Aggregation with Quick Selections (FAQS)} is a new FIB aggregation algorithm, that leverages compact data structures and three unique optimization techniques to quickly and incrementally select next hops when handling route updates. As a result, \textit{FAQS} can run up to 2.53 and 1.75 times faster for IPv4 and IPv6, respectively, than the optimal FIB aggregation algorithm while achieving near-optimal aggregation ratio. Meanwhile, it consumes much less memory and generates much smaller number of FIB changes when carrying out frequent updates. The performance enhancement of the new algorithm addresses many concerns from ISPs regarding performance issues, and enhances the probability to push FIB aggregation techniques further to the level of production adoption by the industry.  
